# Introducing Anisotropic Minkowski Functionals and Quantitative Anisotropy Measures for Local Structure Analysis in Biomedical Imaging

Axel Wismüller[*,1], Titas De[2], Eva Lochmüller[3], Felix Eckstein[3] and Mahesh B. Nagarajan[*,1]

[1]Departments of Biomedical Engineering & Imaging Sciences, University of Rochester, USA
[2]Department of Electrical & Computer Engineering, University of Rochester, USA
[3]Institute of Anatomy and Musculoskeletal Research, Paracelsus Medical University Salzburg, Salzburg, Austria

## ABSTRACT

The ability of Minkowski Functionals to characterize local structure in different biological tissue types has been demonstrated in a variety of medical image processing tasks. We introduce *anisotropic* Minkowski Functionals (AMFs) as a novel variant that captures the inherent anisotropy of the underlying gray-level structures. To quantify the anisotropy characterized by our approach, we further introduce a method to compute a quantitative measure motivated by a technique utilized in MR diffusion tensor imaging, namely fractional anisotropy. We showcase the applicability of our method in the research context of characterizing the local structure properties of trabecular bone micro-architecture in the proximal femur as visualized on multi-detector CT. To this end, AMFs were computed locally for each pixel of ROIs extracted from the head, neck and trochanter regions. Fractional anisotropy was then used to quantify the local anisotropy of the trabecular structures found in these ROIs and to compare its distribution in different anatomical regions. Our results suggest a significantly greater concentration of anisotropic trabecular structures in the head and neck regions when compared to the trochanter region ($p < 10^{-4}$). We also evaluated the ability of such AMFs to predict bone strength in the femoral head of proximal femur specimens obtained from 50 donors. Our results suggest that such AMFs, when used in conjunction with multi-regression models, can outperform more conventional features such as BMD in predicting failure load. We conclude that such anisotropic Minkowski Functionals can capture valuable information regarding directional attributes of local structure, which may be useful in a wide scope of biomedical imaging applications.

## 1. MOTIVATION/PURPOSE

Advanced methods for characterizing local structure properties have been increasingly used in image processing in recent years. Classical methods for texture analysis, such as Gray Level Co-occurrence Matrices (GLCM) [1-2] or wavelet methods [3], have increasingly be complemented by alternative local structure descriptors that exploit local geometric behavior or topological properties for characterizing local structure in pattern recognition problems. Such methods are usually applied in such a manner that structure-characterizing quantities are calculated from image patches or regions of interest (ROIs), where the size and/or shape of such patches is usually confined to simple pre-specified settings in order to enforce locality of feature extraction. Many pattern recognition problems, however, imply rotation-invariant properties, making it desirable for structure descriptors to capture directional preferences in such imaging datasets, which in this body of work will be referred to as capturing "anisotropy". In addition to identifying directional preferences of image features, it is often helpful to also define measures that objectively quantify the degree of rotation-invariance provided by local feature extraction.

*axel.wismueller@gmail.com; phone 585-273-1689; University of Rochester, NY

In this contribution, we introduce a method to extend the capability of Minkowski Functionals (MFs) [4] to capture anisotropic properties in image data. MFs have recently attracted significant attention in a wide scope of pattern recognition domains, including biomedical imaging applications [5-8]. The computation of such measures requires ROIs to be chosen as fixed local patches with sizes determined by the practical applicability within the image processing task at hand, e.g. as (hyper-) spheres or (hyper-) cubes, such as squares in 2D or cubes in 3D. To address the challenge of capturing rotation-invariant image features, we generalize the concept of MFs by introducing so-called Anisotropic Minkowski Functionals (AMFs). This is accomplished by replacing above mentioned naïve ROI definitions with arbitrary kernel functions that will allow us to identify local preferential feature directions in image data. To quantify the degree of anisotropy measured in analyzed ROIs by our approach, we adapt a fractional anisotropy measure motivated by MR diffusion tensor imaging analysis.

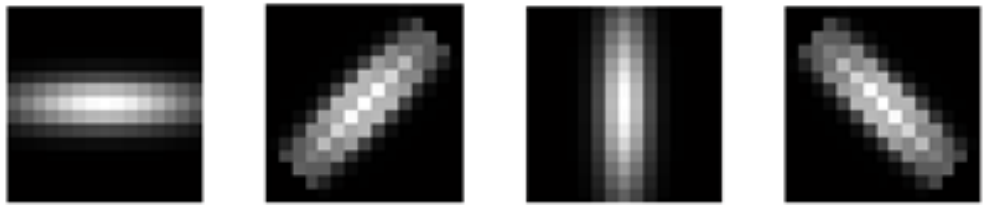

**Figure 1**: Gaussians kernels skewed in 0°, 45°, 90° and 135° (from left to right) used for computation of AMFs.

We demonstrate the applicability of our approach to exploring the degree of anisotropy in trabecular bone structures distributed in different locations of the proximal femur visualized on multi-detector CT. Previous research has revealed that the distribution of trabecular bone is heterogeneous and that structures are anisotropic, i.e., are formed in preferential directions [9], and that the degree of anisotropy varies between different bones, and between different locations within the bones [10]. We investigate the applicability of anisotropic Minkowski Functionals in quantifying the degree of anisotropy in different regions of the femur, i.e. head, neck and trochanter regions for which deviating degrees of anisotropy have been demonstrated in the femoral neck and trochanter, using µCT [10]. We also investigate the use of such AMFs in predicting the bone strength of proximal femur specimens [11]. Previous research has revealed that the bone mineral density of the trabecular compartment in the femoral head is correlated to the failure load of such femur specimens [12]. We compare the ability of our AMF approach to predict the failure load of such femur specimens to that of the conventionally used mean BMD of the femur, as discussed in the following sections. This work is embedded in our group's endeavor to expedite 'big data' analysis in biomedical imaging by means of advanced pattern recognition and machine learning methods for computational radiology, e.g. [13-29].

## 2. DATA

Femur Specimens: 50 left femoral specimens were harvested from fixed human cadavers available from the Institute of Anatomy in Munich. The donors had previously given agreement for their body to be used for purposes of teaching and research. The surrounding soft tissue was excised prior to imaging and failure load testing. The specimens were degassed and sealed in plastic bags filled with 4% formalin/water solution.

Multi-detector Computed Tomography (MDCT): Cross-sectional images of the femora were acquired using a 16-row multi-detector (MD)-CT scanner (Sensation 16; Siemens Medical Solutions, Erlangen, Germany). The specimens were positioned in the scanner as in an in-vivo exam of the pelvis and proximal femur with mild internal rotation of the femur. Each specimen was scanned with a protocol using a collimation and a table feed of 0.75 mm and a reconstruction index of 0.5 mm. A high resolution reconstruction algorithm (kernel U70u) was used, resulting in an in-plane resolution of 0.29 x 0.29 mm$^2$. The image matrix was 512 x 512 pixels, with a field of view of 100 mm. Axial images were acquired where each pixel had dimensions 0.1953x0.1953 mm$^2$ and inter-slice distance was 0.5 mm. Bilinear interpolation was used to create coronal reconstructions from the axial data.

Volume of Interest (VOI) Selection: The outer surface of the cortical shell of the femur was segmented by using bone attenuations of the phantom in each image. A sphere was fitted to the superior surface points of the femoral head using a Gaussian Newton Least Squares technique. The fitted sphere was scaled down to 75% of its original size to account for cortical bone and shape irregularities, and then saved as the femoral head volume of interest (VOI). Further details regarding the fitting algorithm can be found in [12].

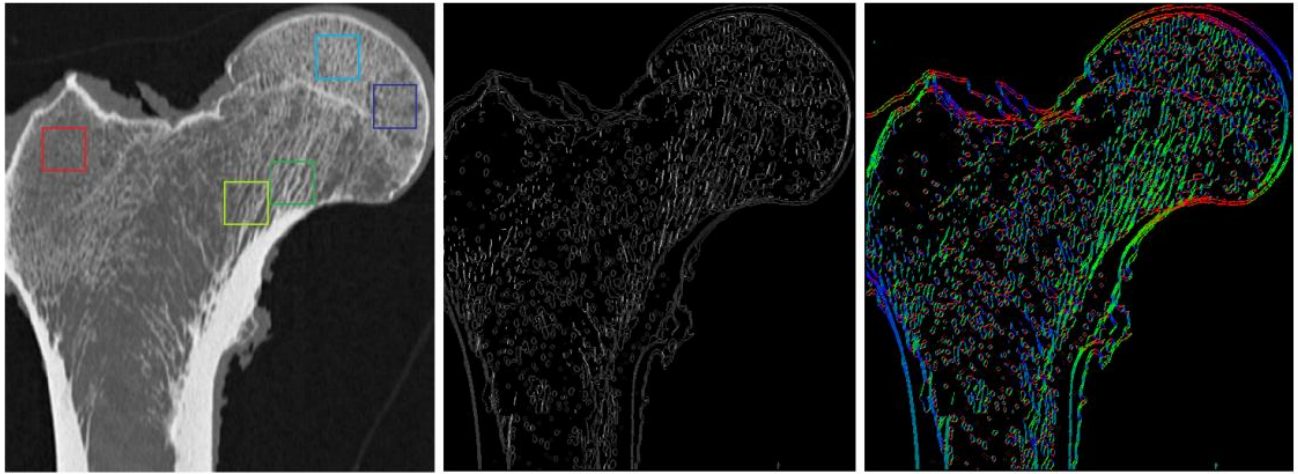

**Figure 2:** LEFT – Coronal reconstruction of femur with ROIs in head (central, light blue), head (medial, dark blue), neck (medial, dark green), neck (lateral, light green) and trochanter (red). MIDDLE – Map of FA values; dark areas correspond to isotropic regions while bright areas correspond to anisotropic regions. RIGHT – Color coded direction map for pixels with FA > 0.03; red corresponds to 0°, green to 60° and blue to 120°.

BMD Measurements: Pixel attenuations in Hounsfield units (HUs) were converted into BMD (mg/cm$^3$) using the calibration phantom with properties, $HA_W$= 0 mg/cm$^3$ and $HA_B$= 200 mg/cm$^3$ (hydroxyapatite for the water-like and the bone-like phase, respectively). If the HU for the water-like and bone-like portions of the phantom are $HU_w$ and $HU_B$ respectively, then one use the previously proposed linear relationship between BMD and HU [11] to calculate the BMD at each voxel as -

$$BMD = [HA_B - HA_w / (HU_B - HU_W)] * (HU - HU_W). \qquad (1)$$

Biomechanical Tests: The failure load was assessed using a side-impact test, simulating a lateral fall on the greater trochanter as described previously [11-12,30]. Briefly, the femoral shaft and head faced downward and the load was applied on the greater trochanter using a universal materials testing machine (Zwick 1445, Ulm, Germany) with a 10 kN force sensor and dedicated software. The failure load was defined as the peak of the load-deformation curve.

## 3. METHODS

### 3.1 Minkowski Functionals

Minkowski Functionals (MF) are used to characterize morphological properties of binary images i.e. shape (geometry) and connectivity (topology) [4]. Three MF features i.e. area, perimeter and Euler characteristic can be calculated from binary images as follows –

$$MF_{area} = n_s, \qquad (2)$$

$$MF_{perimeter} = -4n_s + 2n_e, \qquad (3)$$

$$MF_{Euler} = n_s - n_e + n_v, \qquad (4)$$

where "$n_s$" is the total number of white pixels, "$n_e$" is the total number of edges and "$n_v$" is the number of vertices. The area feature records the number of white pixels in the binary image, the perimeter measure calculates the length of the boundary of white pixel areas and the Euler characteristic is a measure of connectivity between the white pixel regions.

### 3.2 Anisotropic Minkowski Functionals

We introduce anisotropy in the computation of Minkowski Functionals through the use of kernels that provide weights for each of the white pixels, edges and vertices. Although any anisotropic kernel function may be chose, we use Gaussians skewed in the 0°, 45°, 90° and 135° directions to provide a directionally-dependent weighting in the computation of Minkowski Functionals. Examples of such skewed Gaussian kernels are shown in Figure 1. In this study, we fixed the ratio of the major-to-minor radii of the skewed Gaussian to 1:4. The weights for the vertices, edges and white pixels are determined as follows – (1) for each vertex, the average weight of surrounding four pixels, (2) for each edge, the average weight of the two pixels on either side, and (3) for each white pixel, the corresponding weight from the kernel. Thus, four anisotropic measures are computed for each Minkowski Functional outlined in section 1.1.

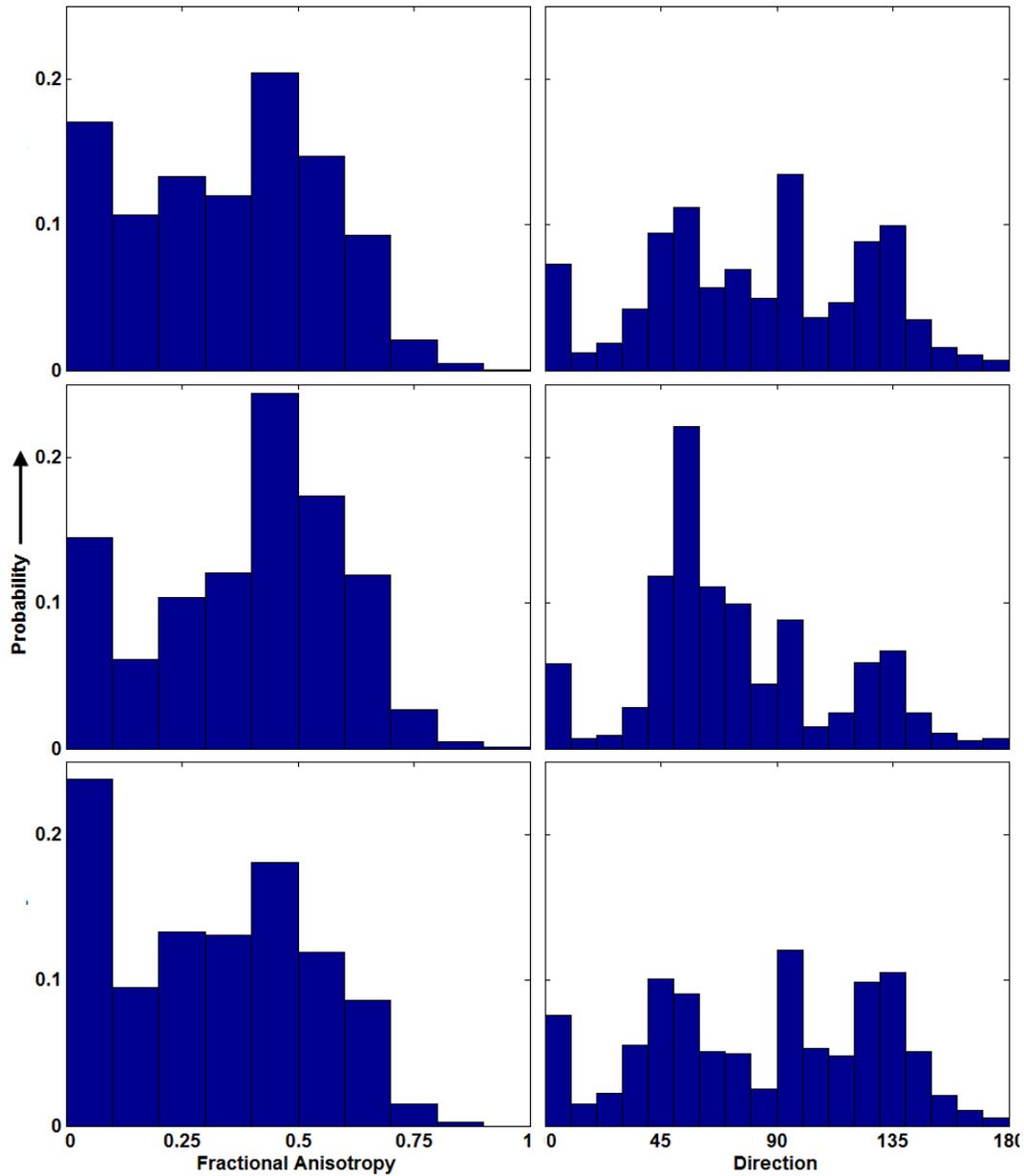

**Figure 3**: LEFT Column – FA histograms for head (top), neck(middle) and trochanter (bottom) regions. RIGHT Column – Direction histograms for head (top), neck (middle) and trochanter (bottom) regions. All histograms were derived from AMF Euler characteristic. Note that the FA histogram for the trochanter region exhibits a greater fraction of isotropic pixels than other regions. The direction histogram for the neck region shows a strong preference for the ~60° direction which is also see on the MDCT image in Fig. 2.

### 3.3 Computation of Anisotropic Minkowski Functionals

Since the ROIs in this study were gray-level images, they were subject to binarization with several thresholds. For each binary image obtained, we computed anisotropic Minkowski Functionals in four directions for a square neighborhood of 5x5 pixels centered on every white pixel. Thus each white pixel is now represented by 4 anisotropic measures for each Minkowski Functional.

### 3.4 Measures of Anisotropy

Once the four directional Minkowski Functionals for a certain measure (eg. Area) are computed for a certain pixel, the magnitude and direction are used to generate 2-D Cartesian coordinates. The four measures are duplicated by mirroring them in the opposite direction to generate a set of eight 2-D coordinates. Principal Component analysis is performed with these eight points to determine the eigenvalues and the corresponding eigenvectors of the point-spread.

We propose to capture the direction of anisotropy by the eigenvector associated with the largest eigenvalue. We further propose to evaluate the anisotropy using a Fractional Anisotropy (FA) measure. For eigenvalues $\lambda_1$ and $\lambda_2$,

$$FA = \frac{\sqrt{(\lambda_1-\lambda_2)^2}}{\sqrt{(\lambda_1)^2 + (\lambda_2)^2}}, \qquad (5)$$

where a value of 0 indicates perfect isotropy while 1 indicates perfect anisotropy in a specific direction. Such an FA measure is computed for each white pixel on every binary image; the FA values for black pixels (background) is set to 0.

In order to combine the FA information computed from all thresholded images, we then assign each pixel the maximum FA value from all the thresholded images. The direction associated with that FA value is also assigned to the pixel. *We* specified an empiric threshold of 0.03 for FA values; pixels with smaller FA were considered to be near isotropic and not assigned a direction. The results of FA and direction assignments are shown in Figure 2.

### 3.5 Quantative Analysis

1. We examined the distribution of anisotropic structures in five ROIs extracted from the following regions of the proximal femur – (1) head (central), (2) head (medial), (3) neck (medial), (4) neck (lateral) and (5) trochanter. Distributions of FA values and angle assignments from different ROIs were converted to distributions of zero mean and unit standard deviation and then compared using a *t*-test.

2. We investigated the use of features derived from AMFs, i.e. histograms of the distribution of FA and angles, extracted from the femoral head of 50 proximal specimens to predict bone strength (failure load). The ability of such features was compared to the previously proposed mean BMD of the femoral head and multi-regression was used to for the machine learning task. Statistical significance was established using the Wilcoxon signed-rank test.

3. We also compared the use of isotropic kernels for extracting Minkowski Functionals to the use of our anisotropic kernels to further analyze the impact of using our approach in characterizing the trabecular bone compartment in the femoral head.

## 4. RESULTS

We observed significant differences between the head, neck and trochanter regions in terms of their FA and direction histograms, as seen in Figure 3. The FA histogram for the trochanter exhibited a much larger fraction of "near-isotropic" pixels (first bin of histogram in Figure 1 left column), i.e. pixels where the structures in the surrounding kernel-defined neighborhood were considered "near-isotropic", while both the head and neck regions exhibited a greater fraction of anisotropic pixels. The direction histogram of the neck region indicated a strong preferential direction among the anisotropic pixels, while the same for the head region displayed a more random distribution of direction.

In terms of bone strength prediction, we noted a significant improvement in performance when using the direction histogram derived from AMF feature Euler Characteristic to characterize the trabecular bone micro-architecture in comparison to using the conventionally used mean BMD ($p < 0.001$). While the best performance was achieved with perimeter-angle, a general improvements in performance was noted with other features derived from AMFs such as perimeter (both FA and direction histogram) and area (FA histogram), as seen in Figure 4.

Finally, we found that the best AMF feature (direction histogram derived from Euler characteristic) also outperforms all MFs extracted with isotropic kernel (square kernel with no Gaussian). However, the feature vectors derived from isotropic MFs also exhibited an improvement in prediction performance over using mean BMD alone ($p < 0.001$), as seen in Figure 4.

## 5. DISCUSSION

While Minkowski Functionals have been previously applied in several medical image processing contexts [5-8], we have proposed a method to extend the capability of such measures to capture anisotropic properties in image data. We accomplish this through the introduction of anisotropic Minkowski Functionals where the ROIs used to compute such measures are defined by arbitrary kernel functions that allow the identification of local preferential feature directions in image data. We also propose a fractional anisotropy measure adapted from MR diffusion tensor imaging to quantify the degree of anisotropy measured in ROIs using such anisotropic Minkowski Functionals. Our method can be extended in multiple ways, e.g. inclusion of 3D anisotropy measures, use of different kernel functions or other anisotropy measures.

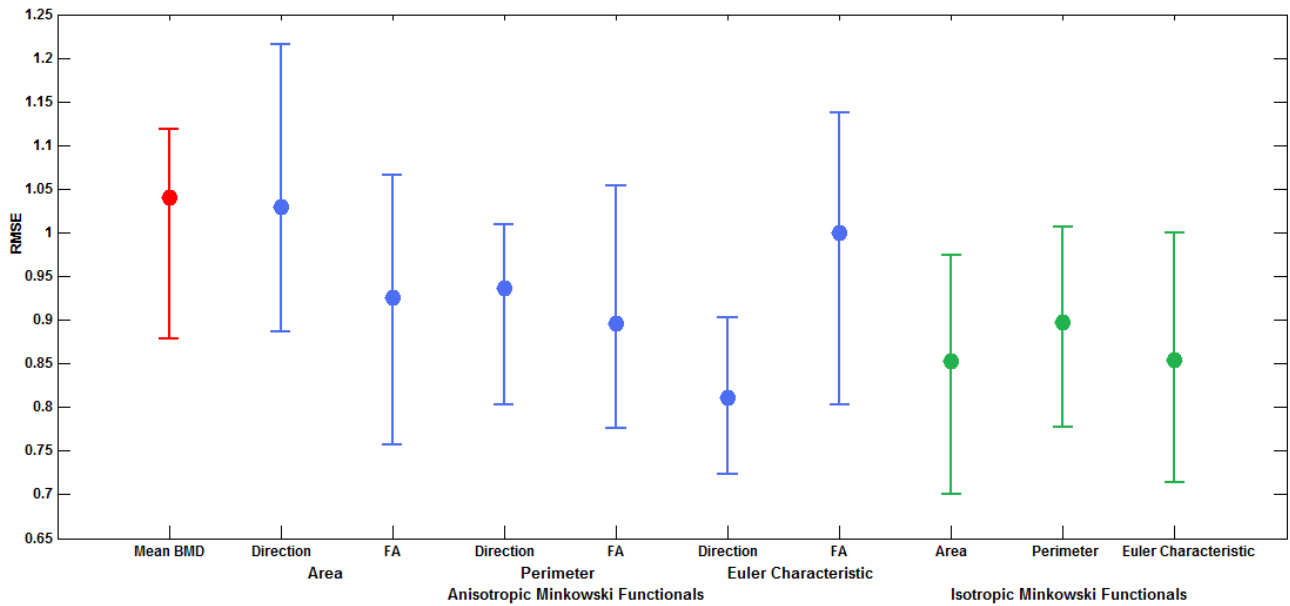

**Figure 4**: Comparison of prediction performance (RMSE) for mean BMD, feature vectors derived from AMFs, and feature vectors derived from isotropic MFs (extracted from the femoral head) when used in conjunction with multi-regression. For each RMSE distribution, the central mark corresponds to the median and the edges are the 25th and 75th percentile. As noted here, all AMF-derived feature vectors outperform the conventionally-used mean BMD. The best performance is achieved with the angle feature vector of AMF perimeter which significantly outperforms all feature vectors derived from isotropic MFs ($p < 0.01$).

We demonstrate the applicability of our approach to exploring the degree of anisotropy in trabecular structures distributed in different portions of the proximal femur visualized on multi-detector CT. Our results suggest that differences in orientation of the trabecular bone micro-architecture in different regions of the proximal femur can be visualized and quantified through such anisotropic Minkowski Functionals. We specifically note that certain findings, such as a high incidence of anisotropic structures in the head and neck regions when compared to the trochanter, and the preferential orientation of structures in the neck when compared to the head region, are in agreement with other studies such as [10,12]. Such an approach has significant potential for quantifying trabecular bone micro-structure and maybe used in addition to conventionally computed density measures such as bone mineral density or bone volume fraction; such measures could serve as diagnostic markers for detection or monitoring of osteoporosis.

We also investigated the use of feature vectors derived from AMFs, i.e. histograms of angles and FA values computed using area, perimeter and Euler characteristic within the femoral head, in their ability to predict bone strength of proximal femur specimens. Our results suggest that such feature vectors, specifically the direction histogram derived from AMF Euler Characteristic, are able to exhibit significant improvements in prediction performance over conventionally used mean BMD ($p < 0.001$), when used in conjunction with multi-regression. This can be attributed to the ability of AMFs to characterize the inherent anisotropy in the trabecular bone micro-architecture; such properties are not adequately captured by mean BMD. We further investigated the use of isotropic MFs, i.e. Minkowski Functionals computed using isotropic square kernels. We do note some improvements in prediction performance over the previously used mean BMD, which suggests that such topological texture feature vectors contribute to an improved characterization of the femoral trabecular bone compartment. However, they still do not account for anisotropy of the structures as evidenced by their poorer performance when compared to AMFs.

## 6. CONCLUSION

This study presents a new approach to computing anisotropic Minkowski Functionals for purposes of capturing anisotropic properties of local tissue structure. We also present a novel approach to quantifying the degree of anisotropy in image data analyzed with our approach through a fractional anisotropy measure. We demonstrate the feasibility of our approach in quantifying the anisotropic properties of trabecular bone extracted from different regions of the proximal femur as imaged by multi-detector CT and show agreement with previous studies that have used μCT. Our results further suggest that features derived from such AMFs can also contribute to improvements in evaluating bone strength and fracture risk in the proximal femur.

## 7. ACKNOWLEDGEMENTS

This research was funded in part by the National Institute of Health (NIH) Award R01-DA-034977, the Harry W. Fischer Award of the University of Rochester, the Clinical and Translational Science Award 5-28527 within the Upstate New York Translational Research Network (UNYTRN) of the Clinical and Translational Science Institute (CTSI), University of Rochester, and by the Center for Emerging and Innovative Sciences (CEIS), a NYSTAR-designated Center for Advanced Technology. The content is solely the responsibility of the authors and does not necessarily represent the official views of the National Institute of Health. We would like to thank M.B. Huber from Department of Imaging Sciences at University of Rochester, Rochester, NY USA, J. Carballido-Gamio, S. Majumdar and T.M. Link from Department of Radiology & Biomedical Imaging at University of California, San Francisco, CA, USA, and J.S. Bauer and T. Baum from Institute of Diagnostic Radiology, Technical University of Munich, Munich, Germany, for their assistance with the data acquisition, VOI annotation, and other intellectual support.

This work is not being and has not been submitted for publication or presentation elsewhere.